
\magnification=1200

\def\ag{{\cal A}/{\cal G}}
\def\agb{\overline{\ag}}
\def\H{{\cal H}_o}
\def\Hd{{\cal H}_d}

\def\tit{\it}
\def\tbf{\bf}
\def\D{{\cal D}}

{}

\def\Comp{{\mathchoice
{\setbox0=\hbox{$\displaystyle\rm C$}\hbox{\hbox to0pt
{\kern0.4\wd0\vrule height0.9\ht0\hss}\box0}}
{\setbox0=\hbox{$\textstyle\rm C$}\hbox{\hbox to0pt
{\kern0.4\wd0\vrule height0.9\ht0\hss}\box0}}
{\setbox0=\hbox{$\scriptstyle\rm C$}\hbox{\hbox to0pt
{\kern0.4\wd0\vrule height0.9\ht0\hss}\box0}}
{\setbox0=\hbox{$\scriptscriptstyle\rm C$}\hbox{\hbox to0pt
{\kern0.4\wd0\vrule height0.9\ht0\hss}\box0}}}}
\def\Co{{\mathchoice
{\setbox0=\hbox{$\displaystyle\rm C$}\hbox{\hbox to0pt
{\kern0.4\wd0\vrule height0.9\ht0\hss}\box0}}
{\setbox0=\hbox{$\textstyle\rm C$}\hbox{\hbox to0pt
{\kern0.4\wd0\vrule height0.9\ht0\hss}\box0}}
{\setbox0=\hbox{$\scriptstyle\rm C$}\hbox{\hbox to0pt
{\kern0.4\wd0\vrule height0.9\ht0\hss}\box0}}
{\setbox0=\hbox{$\scriptscriptstyle\rm C$}\hbox{\hbox to0pt
{\kern0.4\wd0\vrule height0.9\ht0\hss}\box0}}}}
\def\Rl{{\mathchoice
{\setbox0=\hbox{$\displaystyle\rm R$}\hbox{\hbox to0pt
{\kern0.4\wd0\vrule height0.9\ht0\hss}\box0}}
{\setbox0=\hbox{$\textstyle\rm R$}\hbox{\hbox to0pt
{\kern0.4\wd0\vrule height0.9\ht0\hss}\box0}}
{\setbox0=\hbox{$\scriptstyle\rm R$}\hbox{\hbox to0pt
{\kern0.4\wd0\vrule height0.9\ht0\hss}\box0}}
{\setbox0=\hbox{$\scriptscriptstyle\rm R$}\hbox{\hbox to0pt
{\kern0.4\wd0\vrule height0.9\ht0\hss}\box0}}}}
\def\Ac{A^\Co}
\def\Fc{F^\Co}

\centerline{\tenbf A Generalized Wick Transform for Gravity}
\baselineskip=22pt
\vglue 0.8cm
\centerline{\tenrm Abhay Ashtekar}
\baselineskip=13pt
\centerline{\tenit Center for Gravitational Physics and Geometry}
\baselineskip=12pt
\centerline{\tenit Physics Department, Penn State, University Park, PA 16802}
\vglue 0.3cm
\vglue 0.8cm
\centerline{\tenrm ABSTRACT}
\vglue 0.3cm
{\rightskip=3pc \leftskip=3pc \tenrm\baselineskip=12pt\noindent {\bf
Abstract} Using a key observation due to Thiemann, a
generalized Wick transform is introduced to map the constraint
functionals of Riemannian general relativity to those of the
Lorentzian theory, including matter sources. This opens up a new
avenue within ``connection-dynamics'' where one can work, throughout,
only with real variables. The resulting quantum theory would then be
free of complicated reality conditions. Ramifications of this development
to the canonical quantization program are discussed.}

\bigskip

This work is motivated by two related but independent
considerations. The primary motivation comes from canonical quantum
gravity. The approach based on self-dual connections has the advantage
that all equations are low order polynomials. However, to ensure that
one recovers real, Lorentzian general relativity one has to impose
rather complicated ``reality conditions'' on the basic canonical
variables.  (See, e.g., [1,2].)  Quantization would be significantly
easier if one could work entirely with real variables and yet have
manageable constraints. Within connection dynamics, this is indeed
possible in the Riemannian (i.e. positive definite) signature because,
in this case, self-dual connections are real [3]. Therefore, one way
to achieve the desired goal would be to try to define a generalized
Wick transform which would map the Riemannian constraint functionals to
the Lorentzian ones. To be useful, the transform has, of course, to be
sufficiently simple. The purpose of this Letter is show that a recent
result of Thiemann's [4] implies that a transform with desired
features exists and, furthermore, fits rather well with recent
developments in quantum connection-dynamics [5-10]. I should emphasize
that the Riemannian theory plays only a mathematical role in our
description.  The philosophy is the same as the one that underlies
exactly soluble models: The physical --Lorentzian-- theory is
complicated but can be tackled by mapping it to a mathematically
simpler theory. It just happens that the simpler theory can be
identified with Riemannian general relativity.

Our secondary motivation comes from the fact that the
availability of such a transform would also be useful in other
approaches to quantum gravity, notably the ones based on path
integrals. In Minkowskian field theories, one often works with an
Euclidean framework based on the Osterwalder-Schrader axioms,
constructs the theory and then recovers the Wightman functions from
the Schwinger ones by a Wick rotation of the time coordinate. This
simple route is not available in theories of gravity. The question
therefore arises if there is a more general transform which will map
the Riemannian action to the Lorentzian.  We will answer this question
affirmatively using the phase space form of the action. However,
whether this result can be used to develop a full fledged
path integral approach is still unclear.

Let us begin by specifying the phase space. Fix an orientable, smooth
3-manifold $\Sigma$. The phase space will consist of pairs,
$({E}^a_i, K_a^i)$, of real fields on $\Sigma$ where ${E}^a_i$ are
the (non-degenerate) triads of density weight one and $K_a^i$, the
conjugate momenta. Thus, the three metric $q_{ab}$ is defined via
${E}^a_i {E}^{bi} = q q^{ab}$ where $q$ is the determinant of
$q_{ab}$ and the extrinsic curvature $K_{ab}$ is defined by $K_{ab} =
(1/\sqrt{q}) K_a^i {E}^c_i q_{bc}$. The Gauss and the vector
constraints:
$$  {\cal G}_i := \epsilon_{ijk}K_a^j {E}^{ak} = 0\quad {\rm and}
    \quad
     {\cal V}_b:= 2{E}^a_i D_{[a} K_{b]}^i   = 0\, , \eqno(1)$$
where $D$ is defined by $D_a {E}^b_i =0$, are independent of the
signature. The scalar (or Hamiltonian) constraint, on the other hand,
does depend on the signature:
$${\cal S}_L = -qR + 2{E}^{[a}_i {E}^{b]}_j K_a^j K_b^i = 0
 \quad {\rm and} \quad
{\cal S}_R = -qR - 2{E}^{[a}_i {E}^{b]}_j K_a^j K_b^i = 0\, ,
\eqno(2)$$
where and $R$ is the scalar curvature of the metric $q_{ab}$ defined
by ${E}^a_i$ and where the sub-scripts $L$ and $R$ stand for
``Lorentzian'' and ``Riemannian''

To pass to connection dynamics, one can make a canonical
transformation [3,1]:
$$({E}^a_i, K_a^i)\, \mapsto \, (A_a^i := \Gamma_a^i + K_a^i,
{E}^a_i)\eqno(3)$$
where $\Gamma_a^i$ is the $SU(2)$ spin connection determined by the
triad ${E}^a_i$. Note that the new configuration variable $A_a^i$ is
{\it real}; it is again an $SU(2)$ connection on $\Sigma$. It is easy
to verify that the Gauss and the vector constraints (1) can be
expressed as:
$$ {\cal G}_i = \D_a {E}^a_i =0 \quad {\rm and} \quad
{\cal V}_b \approx {E}^a_i F_{ab}^i = 0 \, ,\eqno(4)$$
where $\D$ is the gauge covariant operator defined by $A$ and where
$\approx$ stands for ``equals, modulo Gauss constraint''.  Note
incidentally that these are the simplest equations one can write down
{\it without reference to background fields}: Among the non-trivial
gauge covariant expressions, the left side of the first equation is
the only one which is at most linear in ${E}$ and $A$ and that of the
second is the only one which is at most linear in ${E}$ and quadratic
in $A$. The next simplest equation one can write is:
$$ {\cal S}'_R := \epsilon^{ijk}{E}^a_i {E}^b_j F_{abk} = 0
\eqno(5)$$
When translated in terms of $({E}, K)$, ${\cal S}'_R$ reduces to
${\cal S}_R$ modulo the Gauss constraint. Note that while the left
sides of (2) are non-polynomial in ${E}$ and $K$ due to the presence
of $R$, the left side of (5) is at worst quadratic in ${E}$ and
quadratic in $A$. These are the primary simplifications of connection
dynamics. Using the close similarity of ${\cal S}_R$ and ${\cal
S}_L$,one can readily translate the Lorentzian scalar constraint to
connection dynamics [11]: $ {\cal S}_L \approx - 2R -\epsilon^{ijk}
{E}^a_i {E}^b_j F_{abk} = 0$. However, due to the presence of $R$ the
equation is again complicated and difficult to deal with in the
quantum theory.

Now, if one uses a {\it complex} connection $\Ac_a{}^i = \Gamma_a^i -
iK_a^i$ in place of the real $A_a^i$, the Lorentzian scalar constraint
does simplify [1]: it takes the same form as (5), i.e., becomes
$${\cal S}_L \approx {\cal S}'_L :=\epsilon^{ijk}{E}^a_i {E}^b_j
\Fc_{abk} = 0\, , \eqno(6) $$
where $\Fc$ is the curvature of $\Ac$. However, now the connection is
complex and, to recover real general relativity, one has to impose
reality conditions which also seem complicated at first
sight. However, recently Thiemann [4] has provided an approach to
incorporate such reality conditions in the quantization of a wide
class of theories, including general relativity. This is achieved
through the introduction of a ``complexifier'' which, in the classical
theory, maps real connections $A$ to complex ones $\Ac$. The resulting
quantum complexifier can be regarded as a non-trivial generalization
of the coherent state transform of [12] and (modulo certain technical
issues that are being investigated) maps the Hilbert space of
square-integrable functions of $A$ to an appropriate Hilbert space of
holomorphic functions of $\Ac$. For the class of theories in which the
Hamiltonian (or the Hamiltonian constraint) is simpler in the
holomorphic representation, Thiemann's complexifier should make
dynamics {\it as well as} reality conditions manageable in the
quantum theory.

In this article, I will restrict myself to general relativity but work
entirely with real connections. The resulting framework seems to be
technically simpler and conceptually more transparent for the case
under consideration. I will also present an extension of the
generalized Wick transform to the case when matter sources are present
and discuss several ramifications of these results.

The idea is to construct a Poisson-bracket preserving automorphism on
the algebra of functions on the phase space which maps the Riemannian
constraints to the Lorentzian ones (modulo constant rescalings). Recall
first that, given a real function $T$ on the classical phase space,
the 1-parameter family of diffeomorphisms generated by its Hamiltonian
vector field $X_T$ induces the map $W(t)$ on the algebra of functions
on the phase space:
$$ f\, \mapsto \, W(t)\circ f =
f +t \{f,\, T\} +{t^2\over 2!} \{\{f,\, T\},\, T\}
+ ... \,\,\,
= \sum_{n=0}^{\infty} {t^n\over n!}\,\{f,\, T\}_n \, ,\eqno(7)$$
where $\{\, ,\,\}$ denotes the Poisson bracket.  For each value of the
real parameter $t$, $W(t)$ preserves the $\star$-algebra structure as
well as the Poisson bracket on the space of functions on the phase
space. Now, if we let $T$ be complex-valued, the vector field $X_T$
becomes complex and no longer generates motions on the phase
space. However, (assuming the series converges) the map $W(t)$ of (4)
continues to be a Poisson bracket-preserving automorphism on the
algebra of complex-valued functions on the phase space (although it no
longer preserves the $\star$-relation). Following [4], let us set
$$T := {i\pi\over 2}\int_\Sigma\, d^3x K_a^i {E}^a_i\, .
\eqno(8)$$
Then, regarding ${E}^a_i$ and $K_a^i$ as (coordinate) functions on the
phase space, and setting $W=W|_{t=1}$, we have
$$ W\circ {E}^a_i = i{E}^a_i \quad {\rm and} \quad W\circ K_a^i =
-i K_a^i\, . \eqno(9)$$
The automorphism property now implies $(W\circ f)({E}, K) = f(i{E},
-iK)$, so that the constraint functions transform via:
$$W\circ{\cal G}_i = {\cal G}_i ;\quad W\circ{\cal V}_b =
{\cal V}_b \quad {\rm and}\quad W\circ {\cal S}_R =  - {\cal S}_L
\, .\eqno(10)$$
Thus, the automorphism $W$ defined by the Thiemann generating function
$T$ maps the Riemannian constraints to the Lorentzian ones. It will
therefore be referred to as a ``generalized Wick transform''.

The quantization strategy for the {\it Lorentzian} theory is then as
follows. Begin with the real phase space ${\cal P}$ of pairs $(A_a^i,
{E}^a_i)$. The classical configuration space is then $\ag$, the
space of connections modulo gauge transformations and the quantum
configuration space is a suitable completion $\agb$ thereof [5]. By
now, integral [5-8] and differential [9] calculus on $\agb$ is
well-developed and we can use it to define the Hilbert space of states
and quantum operators. The heuristic requirement that the
configuration and momentum operators $\hat{A}$ and $\hat{E}$ (when
expressed as usual by multiplication by $A$ and $-i\hbar \delta/\delta
A$ respectively) be self-adjoint can be made precise and essentially
suffices to select an unique measure $\mu_o$ on $\agb$.  The resulting
Hilbert space $\H := L^2(\agb ,d\mu_o)$ serves as the space of
kinematic states of quantum gravity, the quantum analog of the {\it
full} phase space ${\cal P}$. (Using integration theory, one can also
define a loop transform from $\H$ to a space of suitable functions of
loops and thus provide a rigorous basis for the Rovelli-Smolin loop
representation [13].) Using differential calculus on $\agb$ , one can
introduce geometric operators on $\H$, e.g., corresponding to areas of
2-surfaces and volumes of 3-dimensional regions [14, 10]. (See also
[15].)  These can be shown to be self-adjoint with {\it purely
discrete} spectra, showing that quantum geometry is very different
from what the continuum picture suggests.

To tackle dynamics, one has to solve the quantum constraints. Since we
work on $\agb$, the Gauss constraint is already taken care
of. (Alternatively, it could also be imposed a la Dirac; the final
result is the same.) The vector or diffeomorphism constraint can be
solved [10] using a ``group averaging'' technique [16]; there are no
anomalies. The space of solutions is naturally endowed with a Hilbert
space structure [10], which we will denote by $\Hd$.

The last and the key step is to solve the Hamiltonian constraint. The
presence of the generalized Wick transform suggests the following
strategy. One can begin with the Riemannian constraint ${\cal S}'_R$
of Eq (15). (Since the Gauss constraint has already been imposed,
${\cal S}_R$ and ${\cal S}'_R$ are on the same footing.)  Because it
has a simple expression in terms of $A$ and $E$, one can hope to
regularize the corresponding quantum operator. For technical reasons
(associated with regularization), one is led to work not with ${\cal
S}'_R$ itself but rather with its positive square-root. Let $\hat{\cal
S}$ be the corresponding quantum operator. The idea now is to exploit
the generalized Wick transform. Since $W$ is a Poisson-bracket
preserving automorphism on the algebra of phase space functions, its
quantum analog $\hat{W}$ would be an automorphism on the algebra of
quantum operators. In view of Eq (10), we can simply {\it define} the
Lorentzian operator $\hat{\cal S}_L$ by:
$$\hat{\cal S}_L = \hat{W}\circ \hat{\cal S}\circ \hat{W}^{-1}
\equiv \exp -{1\over{i\hbar}}\hat {T}\circ \hat{\cal S}\circ
\exp{1\over{i\hbar}}\hat{T}\, ,\eqno(11)$$
where $\hat{T}$ is an operator version of $T$. Eq (7) implies that
$\hat{\cal S}_L$ so defined will automatically have the correct the
classical limit. Physical quantum states $|\Psi\!>_L$ can now be
obtained by Wick transforming the kernel of $\hat{\cal S}$:
$$ \hat{\cal S}|\Psi\!> = 0 \,\,\Leftrightarrow \,\, \hat{\cal S}_L
(\hat W |\Psi\!>) \equiv \hat{\cal S}_L |\Psi\!>_L = 0\,  .\eqno(12)$$
Thus, the availability of the Wick transform could provide
considerable technical simplification: the problem of finding
solutions to all quantum constraints is reduced essentially to that of
regulating relatively simple operators $\hat{\cal S}$ and $\hat{T}$.

Indeed, significant progress has already been made on both these
problems. First, using a key idea due to Rovelli and Smolin [17], the
operator $\hat{\cal S}$ has been made well-defined on diffeomorphism
invariant states, i.e., on a dense subspace of $\Hd$ [18]. (Since
${\cal S}'_R$ itself is only diffeomorphism covariant, the image of
the $\hat{\cal S}$ is also only diffeomorphism covariant. However, we
are interested only in the kernel of this operator.) This
regularization is not yet fully satisfactory because, e.g., it is
``state dependent.''  Nonetheless, it holds considerable promise;
it is the first systematic attempt at a non-perturbative
regularization of the ``Wheeler-DeWitt equation.''  As for $\hat{T}$,
note first that the classical $T$ can be expressed as $T =
(i\pi/2)\{V, H_E\}$ where $V$ is the total volume of $\Sigma$ and $H_E
:= \int\, d^3x\, (1/\sqrt{q}) {\cal S}'_E$ is the ``Riemannian
Hamiltonian''. Hence, it is natural to set: $\hat{T} =
(1/i\hbar)[\hat{V}, \hat{H}_E]$.  Now, $\hat{V}$ has already been
regularized rigorously and the regularization of $\hat{\cal S}$
provides an avenue to regularize $\hat{H}_E$. If this last
regularization can be completed, one would be able to extract
solutions to all quantum constraints via Eq (12).

The final step in the program is to introduce the appropriate inner
product on the space of physical states. If one uses the analog of the
generalized Wick transform for simple model systems one finds that, to
obtain interesting physical states, one has to allow solutions
$|\Psi\!>$ (to the analog of the Riemannian constraint) which are far
from being `tame'; for example, they may diverge at `infinity' (i.e.,
at the boundary of the configuration space). Therefore, the problem of
finding the correct inner product is, in general, quite non-trivial.
However, if these concrete steps can be completed in the case under
consideration, one would have a consistent non-perturbative
quantization of general relativity. The focus will shift to developing
approximation methods to extract physical predictions of the theory.

The `real' strategy adopted here is of course closely related to the
`complex' strategy of Thiemann's [4]. At the classical level, the two
are completely equivalent; only the emphasis is different. Thus, in
the complex approach, one notes that the generalized Wick transform
$W$ has the action $W\circ A_a^i = \Ac_a{}^i$ on connections and
concludes that $W$ sends Riemannian scalar constraint ${\cal S}'_R$ of
Eq (5) to the Lorentzian ${\cal S}'_L$ of Eq (6). Since $\Ac$ and
${\cal S}'_L$ are complex-valued, in the quantum theory, one is then
naturally led to the holomorphic representation. In the real approach,
by contrast, one works exclusively with real phase space variables and
real constraint functions. (In particular, the classical Wick
transform could be useful also in geometrodynamics.) In the quantum
theory, the use of holomorphic representation is no longer
essential. However, there is nothing that prevents one from
constructing this representation using techniques from [4]. Indeed, it
is desirable to construct it because of its closeness to coherent
states; it could for example, play an important role in semi-classical
considerations. In both approaches, the issue of introducing the
physically appropriate inner product remains open, although a general
direction for completing this task has been suggested in [4].

For general relativity, Thiemann introduced the generating function
$T$ only in the source-free case. Can the strategy of
using a generalized Wick transform be extended consistently to
incorporate the presence of matter? The answer is in the affirmative.
Perhaps the most concise way to see this is to use the phase space
action functional. Since this is in part a restatement of the main
results, for the convenience of readers who may be more familiar with
the ADM framework, I will use this opportunity to state these results
using geometrodynamical variables.

The space-time action for general relativity with a cosmological
constant, coupled with a scalar and a Maxwell field, can be expressed
in terms phase space space variables. Modulo surface terms, one obtains:
$$\eqalign{ S^L_R = {1\over 2}\int d^4x\,\, N \big[& \mp (q_{ab}q_{cd}
- q_{ac}q_{bd} - q_{ad}q_{bc})P^{ab}P^{cd}+ 2q(R-2\Lambda) \cr & +
(\mp P^aP^b + B^aB^b)q_{ab} + (\mp \pi^2 + qq^{ab}D_aD_b\phi +\mu^2
q\phi)\big]\cr}\eqno(13a)$$
where the superscript $L$ and the subscript $R$ stand for `Lorentzian'
and `Riemannian' respectively; $q_{ab}$ is the 3-metric and $P^{ab}$
its canonical momentum; $A_a$ is the Maxwell 3-potential and $P^a$ its
canonical conjugate momentum; $\phi$ is the Klein-Gordon field and
$\tilde{\pi}$ its canonically conjugate momentum; $N$ is the lapse;
$R$, the scalar curvature of $q_{ab}$, and, $\Lambda$, the
cosmological constant. To see the explicit form of the constraints,
one can re-express the action in the canonical phase space form as:
$$ S^L_R = \int dt \int d^3x\, P^{ab}\dot{q}_{ab} + P^a \dot{A}_a +
\pi \phi + N{\bf S}^L_R+ 2 N^a({\bf V}^L_R)_a +({}^4\!A\cdot t)
D_aP^a\, ,
\eqno(13b)$$
where $N ,N^a$ and ${}^4\!A$ are the Lagrange multipliers representing
the lapse, the shift and the Maxwell scalar potential; and ${\bf S}$
and ${\bf V}_a$ are the scalar and vector constraints. These are given
by:
$$\eqalign{ 2{\bf S}^L_R =& \pm (q_{ab}q_{cd} - q_{ac} q_{bd} - q_{ad}
q_{bc}) P^{ab} P^{cd} + 2 q(R-2 \Lambda)\cr & + (\pm P^aP^b +
B^aB^b)q_{ab} + (\pm \pi^2 + q q^{ab}D_a\phi D_b\phi + \mu^2q
\phi^2)\cr}
\eqno(14)$$
and
$$2 ({\bf V}^L_R)_a  = 2q_{ab}\,D_cP^bc - \pi D_a\phi - P^b F_{ab}\, ,
\eqno(15)$$
where, $B^a = {\textstyle{1\over2}} \eta^{abc} \partial_{[b}A_{c]}$ is
the magnetic field of the vector potential $A_a$. Note that the scalar
constraint is a density of weight two and the vector potential, of
weight one. Therefore, {\it the lapse} $N$ in Eqs (13) {\it is a scalar
density of weight} $-1$ while the shift $N^a$ is just a vector field.

Now, let us consider the generalized Wick transform ${\bf W}$ on the
Einstein-Maxwell-Klein-Gordon phase space generated by the function
${\bf T}$
$${\bf T} := {i\pi\over 2} \int d^3x\,\, q_{ab}P^{ab}\,\, +\,\,
{i\pi\over 4}\int d^3x\, A_a P^a\, .\eqno(16)$$
It is then straightforward to compute the action of ${\bf W}$ on the
canonical pairs, regarded as (coordinate) functions on the phase
space.  However, since the action functionals of Eq (13) depend also
on the Lagrange multipliers and coupling constants, we need to specify
how ${\bf W}$ acts on them. Can we choose transformation properties of
these non-dynamical variables so that Riemannian action functional is
mapped to the Lorentzian one? Not only does such a choice exist but it
is in fact unique:
$${\bf W}\circ (N, N^a, {}^4\!A\cdot t) = (-N, N^a, e^{i\pi\over 4}\,\,
{}^4\!A\cdot t) \quad{\rm and}\quad {\bf W}\circ (\Lambda, \mu^2)
= (-i\Lambda, -i \mu^2)\, . \eqno(17)$$
(Note that ${\bf W}$ has the same action on the Lagrange multiplier
${}^4\!A\cdot t$ as it has on the dynamical variable $A$.) With this
specification, it is straightforward to verify that: ${\bf W}\circ S_R
= S_L$.  Thus, ${\bf W}$ serves as the generalized Wick
transform. General considerations outlined in the source-free case
suggest that the corresponding quantum operator $\hat{\bf W}$ should
send the kernel of the Riemannian constraint operators to that of the
Lorentzian constraints. However, to make these heuristic
considerations precise, it is essential to regulate the Riemannian
constraint operators and the generator ${\bf T}$. These problems are
yet to be investigated.  Finally, note that the classical generator
${\bf T}$ has a suggestive form:
$$ {\bf T} = \sum_{s=o}^{s=2} ({i\pi\over 4})\times s \int d^3x\,
Q\circ P \, ,\eqno(18)$$
where $s$ is the spin of the field, $Q$ its configuration variable and
$P$ its momentum variable. This form continues to hold for spin
$\textstyle{1\over 2}$ fields as well. It may well be a reflection of
a deeper structure underlying the generalized Wick transform.

To conclude, let me summarize a few features of the classical
generalized Wick transform ${\bf W}$. The fact that ${\bf W}$ sends
the Riemannian action functional to the Lorentzian one may tempt one
to look for a simple space-time interpretation of the
transform. However, I believe that such an interpretation does not
exist. Note in particular that the lapse-shift pairs transform in a
way that is different from what a space-time interpretation would
suggest (i.e., from the common usage in quantum cosmology). The
natural home for the transform appears, rather, to be the phase space.
However, care is needed even in this picture: As was already
emphasized, because the generating function ${\bf T}$ is imaginary,
${\bf W}$ does not arise from a canonical transform on the {\it real}
phase space. We could complexify the phase space and consider the
Hamiltonian flow generated by ${\bf T}$. As far as I can see, however,
one can not interpret ${\bf W}$ as mapping a real subspace of this
complex phase space which can be called ``the phase space of the
Riemannian theory'' to a real subspace which can be identified with
the ``phase space of the Lorentzian theory''. Thus, the interpretation
of $W$ as a generalized Wick transform refers only to its role as an
{\it automorphism on the algebra of functions} on the common real
phase space of the two theories.

Next, while the quantum operator $\hat{\bf W}$ sends solutions of the
Riemannian quantum constraints to solutions of the Lorentzian quantum
constraints, there is no obvious sense in which the classical ${\bf
W}$ maps solutions to constraints of one theory to those of the other
again because ${\bf W}$ is not associated with a diffeomorphism of the
phase space. However, the classical ${\bf W}$ does have an interesting
`dynamical' role. Fix a lapse shift pair $(N, N^a)$ and consider the
`Hamiltonian' functional $H_R := \int d^3x (N{\bf S}_R + N^a({\bf
V}_R)_b)$ of the Riemannian theory.  Denote the corresponding
Hamiltonian vector field by $X^R_{N,{\vec N}}$.  Since the vector
fields can be regarded as derivations on the ring of smooth functions,
the automorphism ${\bf W}$ on the algebra of smooth functions induces
a map on the space of vector fields which we will denote again by
${\bf W}$. Now, because ${\bf W}$ sends $H_R$ to $H_L := \int d^3x
(-N{\bf S}_L + N^a({\bf V}_L)_a)$ and because it preserves Poison
brackets, it follows that ${\bf W}\circ X^R_{N, {\vec N}}= X^L_{-N,
{\vec N}}$. Thus, the Riemannian dynamical trajectories are sent to
the Lorentzian dynamical trajectories. Recall however that only the
integral curves of the Hamiltonian vector fields which lie on the
constraint surface can be identified with physical solutions. Hence,
in general ${\bf W}$ does {\it not} send 4-dimensional Riemannian
solutions to 4-dimensional Lorentzian ones.  On general grounds, one
does not expect any map with this stronger property to exist on the
full solution space. Indeed it is surprising that even a map that
sends $X^R_{N, {\vec N}}$ to $X^L_{-N, {\vec N}}$ should exist. That
this is achieved by an explicit and relatively simple generator ${\bf
T}$ is very striking.  It is quite possible that this fact will have
some powerful applications already in classical gravity.

\vglue 0.6cm
{\bf Acknowledgements}
I am grateful to Thomas Thiemann for sharing his results prior to
publication. I have also benefited from discussions with Jerzy
Lewandowski, Don Marolf and Jose Mour\~ao. This research was supported
in part by the NSF grant PHY93-96246 and by the Eberly research fund
of Penn State University.

\vglue 0.6cm
{\bf 6. References}
\vglue 0.4cm

\itemitem{1.} A. Ashtekar, {\tit Phys. Rev. Lett.} {\tbf 57} (1986) 2244;
{\tit Phys. Rev.} {\tbf D36} (1987) 1587.
\itemitem{2.}A. Ashtekar, {\tit Lectures on Non-Perturbative Canonical
Gravity}, Notes prepared in collaboration with R. S. Tate (World
Scientific, Singapore, 1991).
\itemitem{3.} A. Ashtekar, in {\tit Mathematics and General Relativity}
edited by J. Isenberg, (AMS, Providence, \- 1987).
\itemitem{4.} T. Thiemann, {\tit Reality conditions inducing transforms for
quantum gauge field theory and quantum gravity}, (pre-print).
\itemitem{5.} A. Ashtekar and C. J. Isham, {\tit Class. and Quantum Grav.}
{\tbf 9} (1992) 1433.
\itemitem{6.} A. Ashtekar and J. Lewandowski, in {\tit Knots and Quantum
Gravity}, ed. J. Baez (Oxford University Press, Oxford, 1994), {\tit
J. Math. Phys.}  {\tbf 36} (1995) 2170.
\itemitem{7.} J. Baez, {\tit Lett. Math. Phys.}{\tbf 31} (1994) 213;
in {\tit The Proceedings of the Conference on Quantum Topology}, ed
D. N. Yetter (World Scientific, Singapore, 1994); J. Baez and
S. Sawin, {\tit Functional integration of spaces of connections},
q-alg/9507023.
\itemitem{8.} D. Marolf and J. Mor\~ao, {\tit Commun. Math. Phys.}
{\tbf 170} (1995) 583.
\itemitem{9.} A. Ashtekar and J. Lewandowski, {\tit J. Geo. \& Phys.}
(in press).
\itemitem{10.} A. Ashtekar, J. Lewandowski, D. Marolf, J. Mour\~ao and
T. Thiemann, {\tit J. Math. Phys.} {\tbf 36} (1995) 6456.
\itemitem{11.}  J.F. Barbero G., {\tit Phys. Rev.} {\tbf D51}
(1995) 5507.
\itemitem{12.} A. Ashtekar, J. Lewandowski, D. Marolf, J. Mour\~ao and
T. Thiemann, {\tit Coherent state transform on the space of connections},
{\tit J. Funct. Analysis} (in press).
\itemitem{13.} R. Rovelli and L. Smolin, {\tit Nucl. Phys.} {\tbf B331}
(1990) 80.
\itemitem{14.} A. Ashtekar and J. Lewandowski, {\tit Quantum Geometry},
(pre-print).
\itemitem{15.} C. Rovelli and L. Smolin, {\tit Nucl. Phys.} {\tbf B442},
(1995) 593; R. Loll, {\tit The volume operator in discretized
gravity}, (pre-print).
\itemitem{16.} A. Higuchi, {\tit Class. \& Quantum Grav.} {\tbf 8} (1991)
1983, 2023; N. P. Landsman, {\tit J. Geo. \& Phys.} {\tbf 15} (1995) 285.
\itemitem{17.} C. Rovelli, {\tit J. Math. Phys} {\tbf 36} (1995).
\itemitem{18.} A. Ashtekar and J. Lewandowski, (in preparation).

\end\bye